\documentclass[reprint,superscriptaddress]{revtex4-1}
\usepackage{graphicx}
\usepackage{amsmath}

\begin{document}

\title{Deep and persistent spectral holes in thulium-doped yttrium orthosilicate for imaging applications}

\author{C. Venet}
\affiliation{Institut Langevin, Ondes et Images, ESPCI ParisTech, PSL Research University, CNRS UMR 7587, INSERM U979, Universit\'e Paris VI Pierre et Marie Curie, 1 rue Jussieu, 75005 Paris, France}
\affiliation{Laboratoire Aim\'e Cotton, CNRS UMR 9188, Univ. Paris-Sud, ENS-Cachan, Universit\'e Paris-Saclay, 91405, Orsay, France}
\author{B. Car}
\affiliation{Laboratoire Aim\'e Cotton, CNRS UMR 9188, Univ. Paris-Sud, ENS-Cachan, Universit\'e Paris-Saclay, 91405, Orsay, France}
\author{L. Veissier}
\affiliation{Laboratoire Aim\'e Cotton, CNRS UMR 9188, Univ. Paris-Sud, ENS-Cachan, Universit\'e Paris-Saclay, 91405, Orsay, France}
\author{F. Ramaz}
\affiliation{Institut Langevin, Ondes et Images, ESPCI ParisTech, PSL Research University, CNRS UMR 7587, INSERM U979, Universit\'e Paris VI Pierre et Marie Curie, 1 rue Jussieu, 75005 Paris, France}
\author{A. Louchet-Chauvet}
\affiliation{Laboratoire Aim\'e Cotton, CNRS UMR 9188, Univ. Paris-Sud, ENS-Cachan, Universit\'e Paris-Saclay, 91405, Orsay, France}
\email[]{anne.louchet-chauvet@u-psud.fr}

\begin{abstract}
With their optical wavelength in the near infrared (790-800nm) and their unique spectroscopic properties at cryogenic temperatures, thulium-doped crystals are at the center of many architectures linked to classical signal processing and quantum information. In this work, we focus on Tm-doped YSO, a compound that was left aside in the mid-1990s due to its rather short optical coherence lifetime. By means of time-resolved hole-burning spectroscopy, we investigate the anisotropic enhanced nuclear Zeeman effect and demonstrate deep, sub-MHz, persistent spectral hole burning under specific magnetic field orientation and magnitude.
By estimating the experimental parameters corresponding to a real-scale ultrasound optical tomography setup using Tm:YSO as a spectral filter, we validate Tm:YSO as a promising compound for medical imaging in the human body.

\end{abstract}

\maketitle



Rare-earth-doped compounds have been attracting growing attention over the past three decades due to the exceptional spectroscopic properties they display at cryogenic temperature. They are characterized by the combination of a broad inhomogeneous absorption line and a narrow homogeneous width, leading to spectral holeburning and photon echo capacity.
These compounds are particularly versatile since there are numerous host-dopant combinations leading to a wide variety of spectroscopic properties~\cite{sun2005book,thiel2011RE}.
First considered as possible materials for optical data storage~\cite{mitsunaga1991opticalmemory}, they became the cornerstone of many signal processing architectures, including spectral analysis and processing of optically-carried radiofrequency signals~\cite{babbitt2014,berger2016}.
The attention is even greater today as they are considered for quantum information storage and quantum technology challenges~\cite{tittel2010,simon2017quantum}.
This huge research effort at the international scale has driven the search for the perfect compound over the last $20$ years, with a strong focus on demonstrating long optical and spin coherence lifetimes~\cite{rancic2018coherence}.

Nevertheless, propositions using rare-earth doped materials have been made recently addressing new applications, including laser stabilization for optical clocks~\cite{thorpe2011} and spectral filtering for \emph{in-vivo} medical imaging~\cite{li2008uot}. In these domains, the performance criteria can strongly differ from that of signal processing or quantum memory architectures. It is therefore appealing to reconsider some rare-earth doped compounds in the light of these new paradigms.

Let us focus on spectral filtering, and more specifically on ultrasound optical tomography, where rare-earth based spectral filters can be used to isolate the signal of interest~\cite{li2008uot}. These spectral filters are particularly relevant for medical purposes because they are intrinsically immune to speckle decorrelation, allowing for \emph{in-vivo} imaging. For these applications, the operation wavelength (determined by the dopant) must be in the optical therapeutic window (650-1000nm) where the absorbance of living tissues is minimized~\cite{richards1996}.
In this wavelength range, only Nd, Yb, Ho and Tm-doped compounds are available. Up to now, the only ultrasound optical tomography demonstrations in the optical therapeutic window using a rare-earth-doped crystal as a spectral filter involve the Tm$^{3+}$ ion with its $^3$H$_6\rightarrow ^3$H$_4$ transition around $790-800$~nm, and more specifically Tm-doped YAG~\cite{li2008uot,venet2018}. Praseodymium-based filters have also been demonstrated~\cite{zhang2012}, but their $606$~nm operation wavelength is not compatible with medical imaging because of the absorbance of hemoglobin.
The choice of Tm:YAG over the many other possible Tm-doped compounds is justified by the abundant literature on this material~\cite[and references therein]{sun2005book}, mainly driven by research towards signal processing and quantum memories. However, for ultrasound optical tomography, coherence lifetimes are not crucial parameters, whereas the contrast and lifetime of the spectral holes are paramount to allow large filtering contrast and maximize the duty cycle dedicated to imaging~\cite{venet2018}.

\begin{figure}[t]
    \centering
    \includegraphics[width=0.95\linewidth]{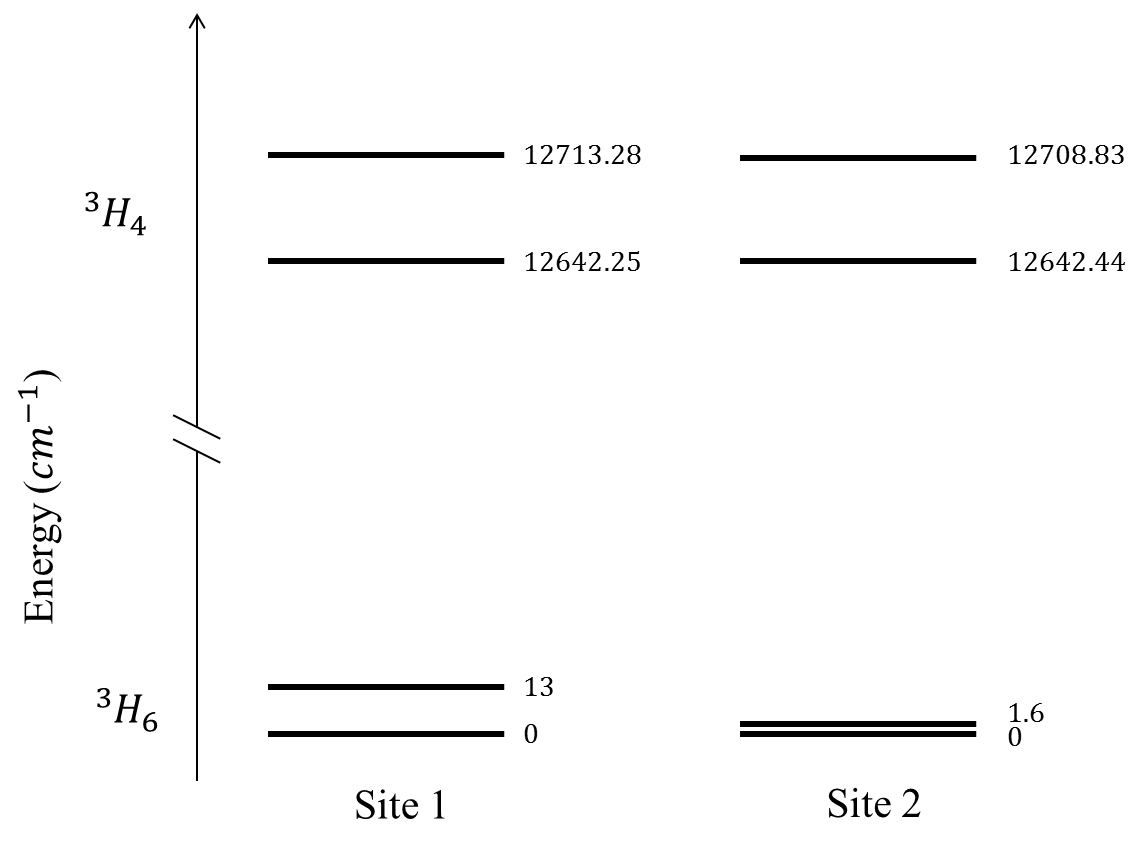}
    \caption{Energy level diagram for sites 1 and 2 of Tm:YSO (as measured by R. Equall~\cite{equallPhD}).}
    \label{fig:niveauxTmYSO}
\end{figure}

Among the other possible candidates (Tm:YSO, Tm:YGG, Tm:LuAG, Tm:LaF$_3$, Tm:LiNbO$_3$,...), Tm:YSO (yttrium orthosilicate Y$_2$SiO$_5$) is the crystal where the optical inhomogeneous linewidth is narrowest~\cite{sun2005book}: $\Gamma_{\textrm{inh}}=2.6$~GHz and $4.5$~GHz for sites 1 and 2, respectively. This line narrowness combined with the low number of crystallographic substitution sites results in a large absorption coefficient. As an example, for a given doping concentration, the absorption coefficient in Tm:YSO is $7$ times larger than in YAG.
This is interesting for ultrasound optical tomography since large filtering dynamics are needed, and a minimal inhomogeneous linewidth of only a few tens of MHz is necessary.
Tm:YSO was left aside from quantum information applied research mainly due to the proximity of the nearest Stark level in the electronic ground state ${}^3H_6$ (around $13$~cm$^{-1}$ and $1.6$~cm$^{-1}$ for sites 1 and 2, respectively) (see Fig.~\ref{fig:niveauxTmYSO}). These closeby Stark levels induce an enhanced nuclear magnetic moment~\cite{abragam1970}, which results in a strong sensitivity to magnetic fluctuations and therefore to low optical coherence lifetimes in both sites ($T_2=4.8~\mu$s and $1.3~\mu$s)~\cite{equallPhD}.
The minimal spectral hole width being given by $2/\pi T_2$, such coherence lifetimes still make $1$~MHz-wide spectral holes possible. Therefore Tm:YSO may still be compatible with ultrasound optical tomography.

Overall, the narrow optical absorption line, the low number of sites, and the MHz-wide spectral holes make Tm:YSO a promising compound for ultrasound optical tomography applications. However, persistent spectral holeburning with high contrast has yet to be demonstrated to consider Tm:YSO as a valid substitute for Tm:YAG for this application.

With its even number of electrons, Tm$^{3+}$ is a non-Kramers ion. When it is placed in a low symmetry environment, the electronic level degeneracy can be totally lifted by the crystal field~\cite{abragam1970}. The only natural isotope $^{169}$Tm exhibits a $I=1/2$ nuclear spin. Therefore the two nuclear states are degenerate in the absence of external magnetic field due to the quenching of the total electronic angular momentum $J$. One way to create a persistent spectral hole in such a system is to apply a magnetic field to split the ground state, and use one of these ground states as a storage level. This has been done successfully in Tm:YAG~\cite{louchet2007branching}. In that respect, Tm strongly differs from all the other non-Kramers rare-earth ions that exhibit a quadrupole structure at zero magnetic field ($I=3/2$ or higher, and particularly $I=5/2$ in Pr$^{3+}$ and both isotopes of Eu$^{3+}$).   

The first section of this paper is devoted to the description of the experimental setup. In the second section, we present absorption and holeburning measurements in Tm:YSO without a magnetic field, and observe transient storage in the metastable state. In the third section, we study the effect of an external magnetic field on the holeburning relaxation dynamics, and evidence deep and persistent spectral holes. Finally we extrapolate the experimental results to evaluate the potential of Tm:YSO for realistic imaging applications.

\section{Experimental details}
The work reported in this paper is based on experimental work using a 0.1\% at. Tm:YSO crystal (Scientific Materials). The crystal is cooled down to $2.1$~K in a liquid helium cryostat from ICE Oxford equipped with AR-coated windows. The crystal is held in helium vapor above the superfluid liquid helium surface. It is mounted on an Attocube ANRv51 rotator. The crystal is cut in such a way that the $b$ axis coincides with the rotation axis and the light propagation direction. Its thickness along the propagation direction is $2$~mm.

The crystal orientation in the laboratory frame can be measured optically by finding the principal axes direction, which gives an angular accuracy of a few degrees. The light polarization is linear and parallel to $D_2$ unless stated otherwise. A vertical homogeneous magnetic field is applied using a 3T superconducting Helmholtz coil pair in the sample space. When the crystal is rotated, the magnetic field spans the $(D_1,D_2)$ plane. The field orientation is defined by the angle $\theta$ that it forms with the $D_1$ axis.

The optical pulse sequence is obtained from a free-running continuous-wave Toptica DLPro laser source emitting at  $12642.25$~cm$^{-1}$ (\emph{ie} $791$~nm). This wavelength specifically addresses the Tm$^{3+}$ ions in site 1. The light is amplified with a Toptica BoosTA tapered amplifier, and shaped with $110~$MHz acousto-optic modulator (AOM, AA Opto-Electronic) driven by an arbitrary waveform generator (Tektronix AWG5004). The beam is loosely focused on a $200~\mu$m diameter spot on the crystal, with a peak power of the order of $50~\mu$W. The transmitted light is collected on an avalanche photodetector (Thorlabs APD110A).

\section{Spectroscopy with no magnetic field}

\subsection{Absorption}
\begin{figure}[t]
\centering\includegraphics[width=0.9\linewidth]{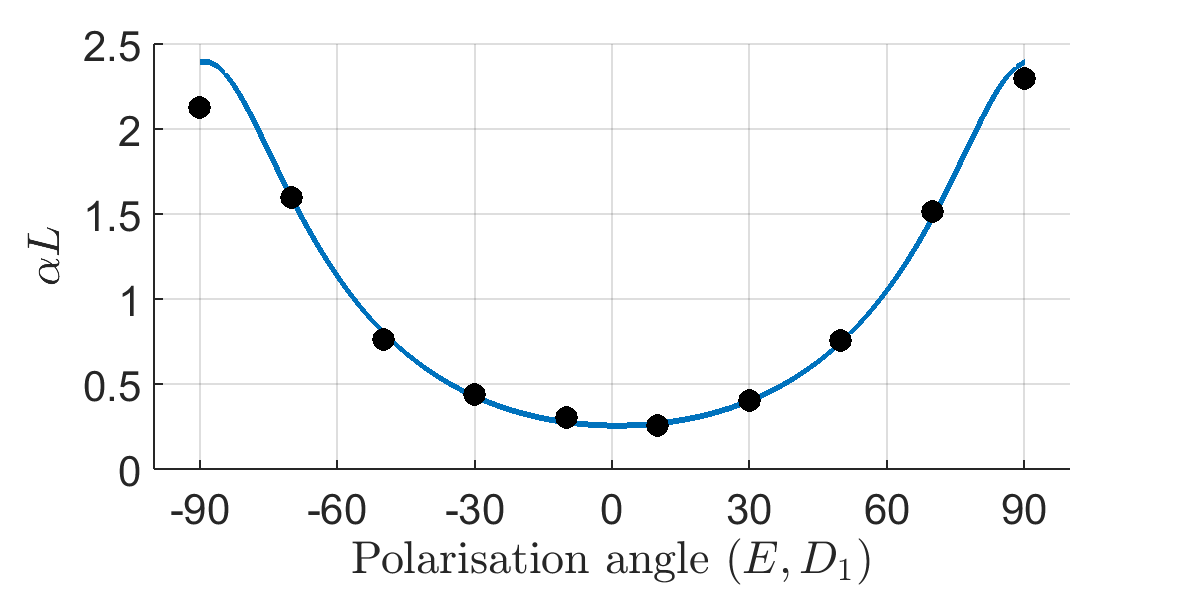}
\caption{Absorption vs polarization}
\label{fig:AbsPolar}
\end{figure}

We measure the optical depth $\alpha L$ along the $^3$H$_6\rightarrow ^3$H$_4$ optical transition for various linear input polarization orientations in the $(D_1,D_2)$ plane by performing an optical nutation experiment~\cite{allen1987} and comparing the transmitted power at the beginning and end of the nutation signal (see Fig.~\ref{fig:AbsPolar}). The crystal absorption varies by a factor of $\sim 10$ when the polarization direction is rotated from $D_1$ to $D_2$. The absorption is maximal when the light is polarized along $D_2$.

In agreement with R. Equall's PhD work~\cite{equallPhD}, we measure an absorption coefficient per unit doping percentage of $110/$cm$/\%$ along $D_2$. It is much larger than in Tm:YAG ($16/$cm$/\%$). The oscillator strength is~\cite{hilborn1982}:
\begin{equation}
    f_{\textrm{YSO}}=1.9\cdot 10^{-8},
\end{equation}
significantly smaller compared to $f_{\textrm{YAG}}=8.3\cdot 10^{-8}$ in YAG, but thanks to the low number of sites (2 sites in YSO, compared to 6 sites in YAG) and the narrow inhomogeneous broadening, quite large opacities are accessible while keeping a low doping concentration and small crystal size.

\subsection{Hole burning}
We set up a holeburning sequence, composed as follows: first the crystal is pumped with a $5$~ms-long monochromatic laser pulse. After a $100~\mu$s waiting time, the hole is read out with an attenuated $400~\mu$s pulse, frequency chirped over $20$~MHz. 

\begin{figure}[t]
\centering\includegraphics[width=0.9\linewidth]{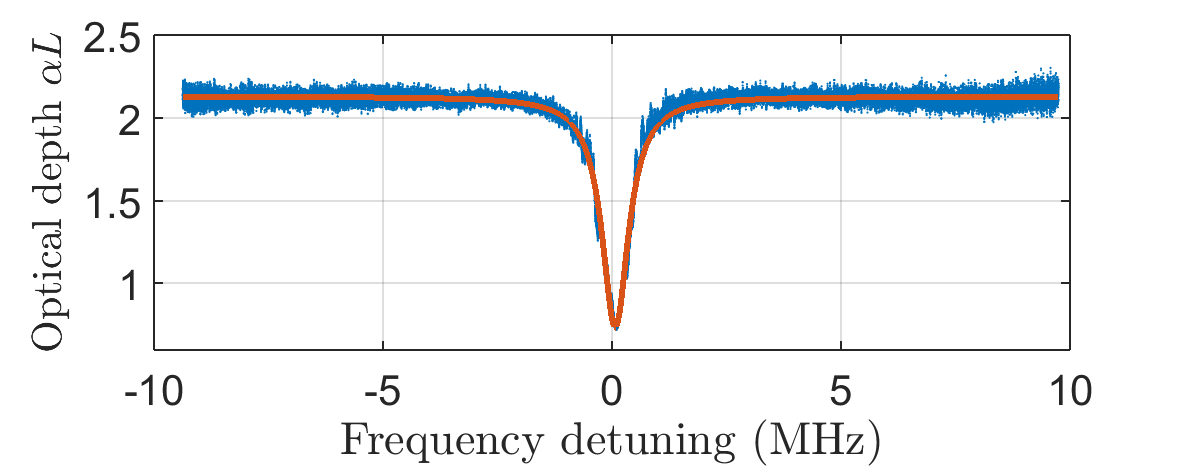}
\caption{Normalized optical depth of Tm:YSO $100~\mu$s after burning a spectral hole, without a magnetic field.  The red line is a Lorentzian function fitted to the data.}
\label{fig:TrouTmYSO_noB}
\end{figure}

Figure~\ref{fig:TrouTmYSO_noB} shows a spectral hole in the absorption profile of Tm:YSO measured with such a pulse sequence. The hole is present although the waiting time is significantly larger than the excited state lifetime ($55~\mu$s according to~\cite{equallPhD}), indicating that the atoms are briefly stored in another state which we assume is the intermediate $^3F_4$ state, often labelled as metastable. The hole shape is lorentzian, and its width is $705$~kHz (FWHM), in excess compared to the limit imposed by the coherence lifetime ($2/(\pi T_2)=133$~kHz) due to saturation broadening.

\begin{figure}[t]
\centering\includegraphics[width=0.9\linewidth]{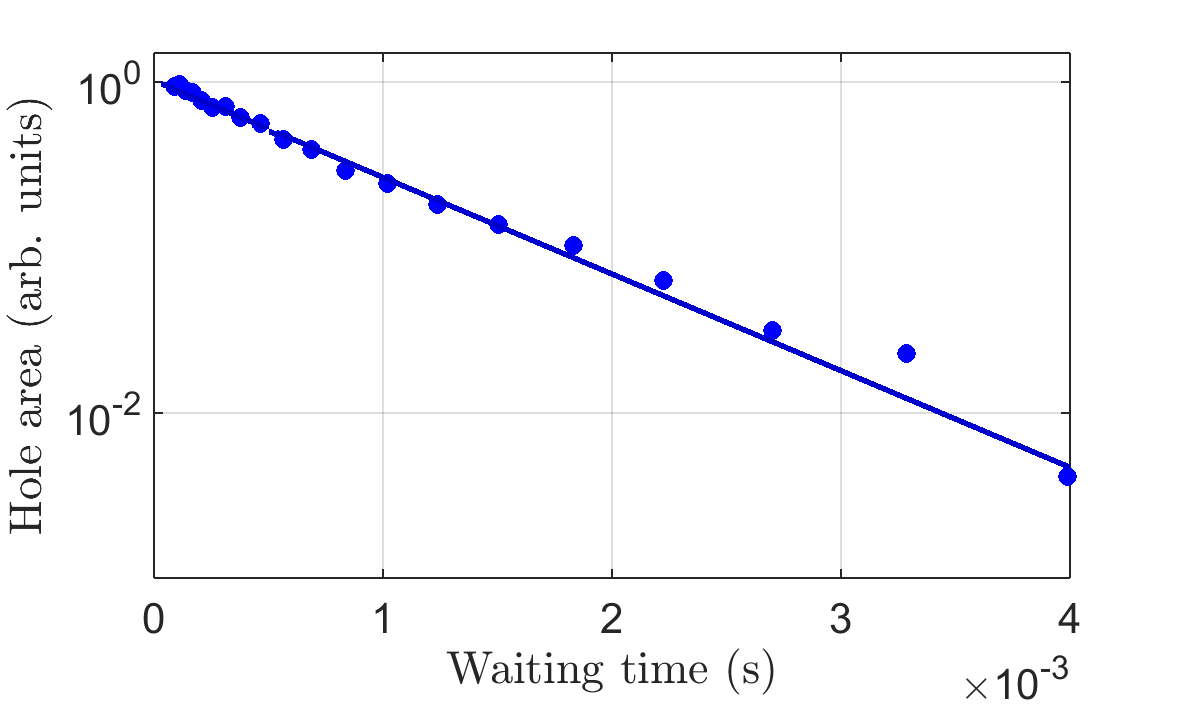}
\caption{Hole burning decay without magnetic field, revealing a $740~\mu$s lifetime. The dots correspond to experimental data, and the line to the exponential decay fit.}
\label{fig:TmYSOLifetimes_noB}
\end{figure}

By varying the waiting time up to $4$~ms, we observe an exponential decay of the hole area (see Fig.~\ref{fig:TmYSOLifetimes_noB}), which we assign to the metastable level population decay $T_m=740~\mu$s. This lifetime is considerably shorter than the fluorescence lifetimes reported in more strongly doped Tm:YSO samples~\cite{li1991fluoTmYSO}, suggesting that the decay is mostly non-radiative. 

\section{Spectral holeburning under magnetic field}
Applying a magnetic field on Tm-doped YSO splits each electronic level into a pair of Zeeman sublevels. We will note these splittings $\Delta_g$ and $\Delta_e$ (expressed in Hz), for the ground and excited state, respectively. We refer to these levels as Zeeman but the splitting mechanism is expected to include other interactions such as the second-order coupling between the electronic Zeeman interaction and the hyperfine interaction, resulting in a so-called \emph{enhanced nuclear Zeeman interaction}. In a low symmetry site such as in YAG, this leads to strongly anisotropic Zeeman splittings in both electronic states $^3H_6$ and $^3H_4$, with maximum sensitivities in the ground state of the order of $400$~MHz/T.

In YSO the substitution sites 1 and 2 have even lower symmetry ($C_1$). We expect a behavior of the Tm ion under magnetic field very similar to that in YAG. Due to the larger Stark splitting in the excited state ($71$~cm$^{-1}$), the enhancement of the nuclear Zeeman effect is expected to be smaller than in the ground state where the Stark splitting is only $13$~cm$^{-1}$.
Some spectroscopic studies under magnetic field in Tm:YSO were reported in~\cite{equallPhD}, but the orientation was not known and the magnetic fields were very large (over $4$~T).

For moderate ground state splittings ($\Delta_g \ll k_B T/\hbar$), the initial population is equally distributed among the two ground state sublevels. Therefore we expect the monochromatic excitation to optically pump Tm ions out of the resonant ground Zeeman sublevel and accumulate them in the non-resonant one, leading to side holes and antiholes. A fraction of ions may still be stored in the metastable state and visible at small delays. Thus the holeburning spectrum evolves in two ways: extra features will appear on each side of the main hole, and this complex structure will exhibit multiple lifetimes.

\begin{figure}[t]
\centering\includegraphics[width=0.9\linewidth]{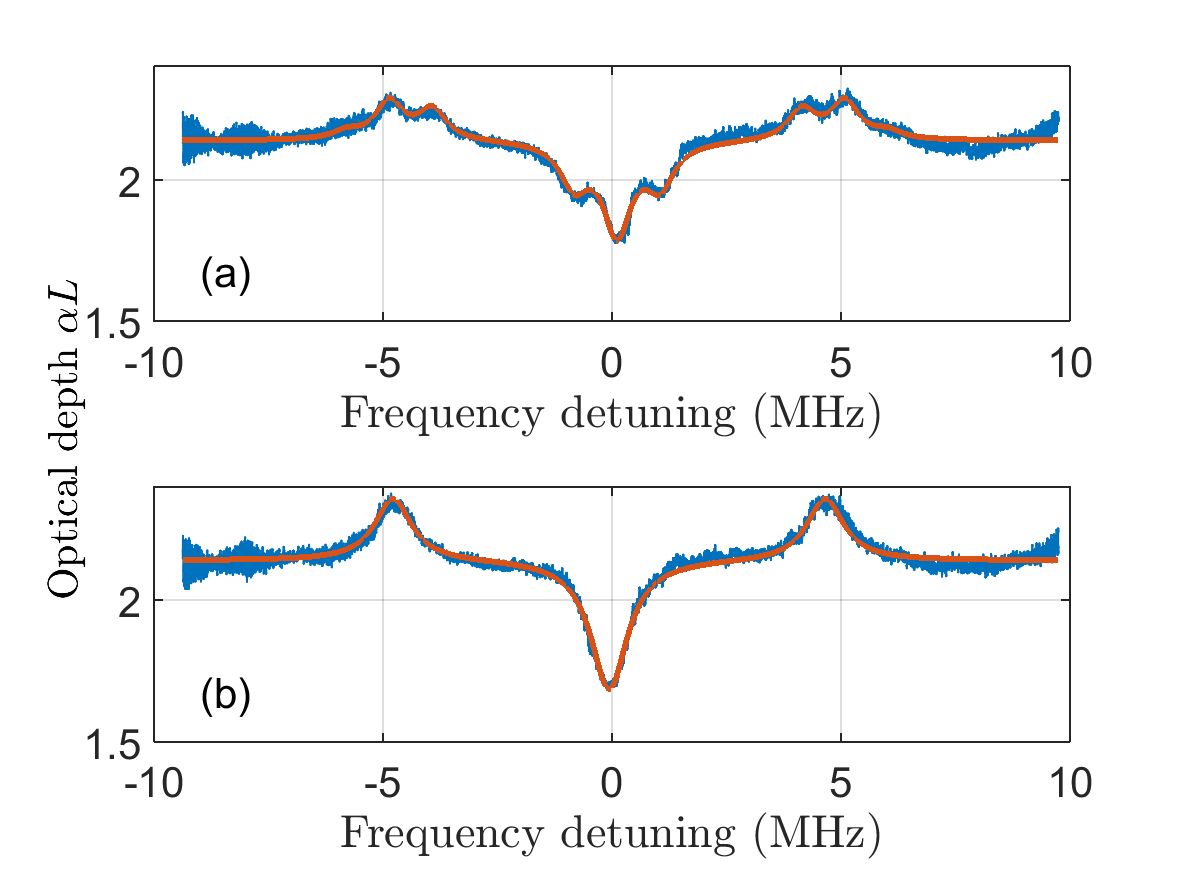}
\caption{Tm:YSO optical depth $2$~ms after burning a spectral hole (a) with a $45$~mT magnetic field $41^\circ$ from the $D_1$ axis, (b) with a $6.6$~mT magnetic field $330^\circ$ from the $D_1$ axis. The red lines are sums of Lorentzian functions to guide the eye.}
\label{fig:TrouTmYSO}
\end{figure}

Figure~\ref{fig:TrouTmYSO} shows holeburning spectra recorded $2$~ms after shining a monochromatic pulse on the crystal, with a magnetic field in two different orientations in the $(D_1,D_2)$ plane: $\theta=41^\circ$ (a) and $\theta=330^\circ$ (b). The positions of the holes and antiholes provide us with a direct measurement of the Zeeman splitting values~\cite{deseze2006}. In both experiments, the magnetic field is adjusted so that all the side structures are visible in the limited chirp range allowed by the AOM.

In Fig.~\ref{fig:TrouTmYSO}(a) we observe two side holes and six antiholes. The side holes are located $\pm \Delta_e$ from the central hole. The six antiholes are located at positions $\pm(\Delta_g-\Delta_e)$,  $\pm(\Delta_g)$ and $\pm(\Delta_g+\Delta_e)$. In Fig.~\ref{fig:TrouTmYSO}(b) the side holes are not resolved and there are only two visible antiholes. As will be shown later, this is due to the small value of $\Delta_e$ with respect to $\Delta_g$, leading to significant overlapping. Figures~\ref{fig:TrouTmYSO}(a) and ~\ref{fig:TrouTmYSO}(b) are fitted with a sum of Lorentzian functions with $745$~kHz and $979$~kHz FWHM, respectively.

Note that the magnetic field in Fig.~\ref{fig:TrouTmYSO}(a) is 7 times stronger than in Fig.~\ref{fig:TrouTmYSO}(b). This anisotropic Zeeman effect is investigated in more detail in the following paragraph.

\subsection{Anisotropy of the enhanced nuclear Zeeman effect}
We perform holeburning spectroscopy for various magnetic field orientations in the $(D_1,D_2)$ plane, by rotating the crystal around its $b$-axis.
The chirp interval being limited by the AOM bandwidth, we adjust the magnetic field in order to keep the side structures visible. Due to the very large discrepancy between the ground and excited state splittings, we perform two independent series of measurements, with a low magnetic field to probe the ground state splitting  and a larger magnetic field to probe the excited state splitting.

\begin{figure}[t!]
\centering\includegraphics[width=0.9\linewidth]{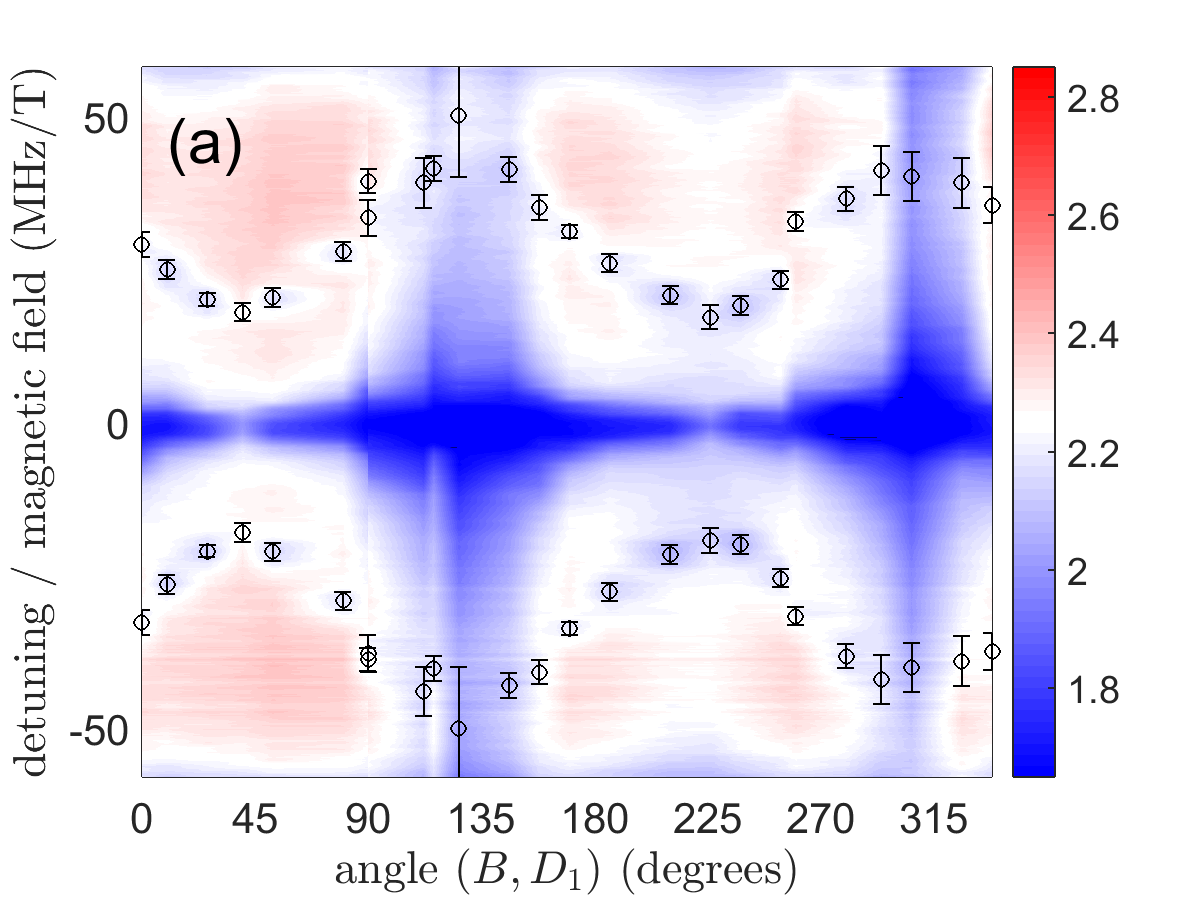}
\centering\includegraphics[width=0.9\linewidth]{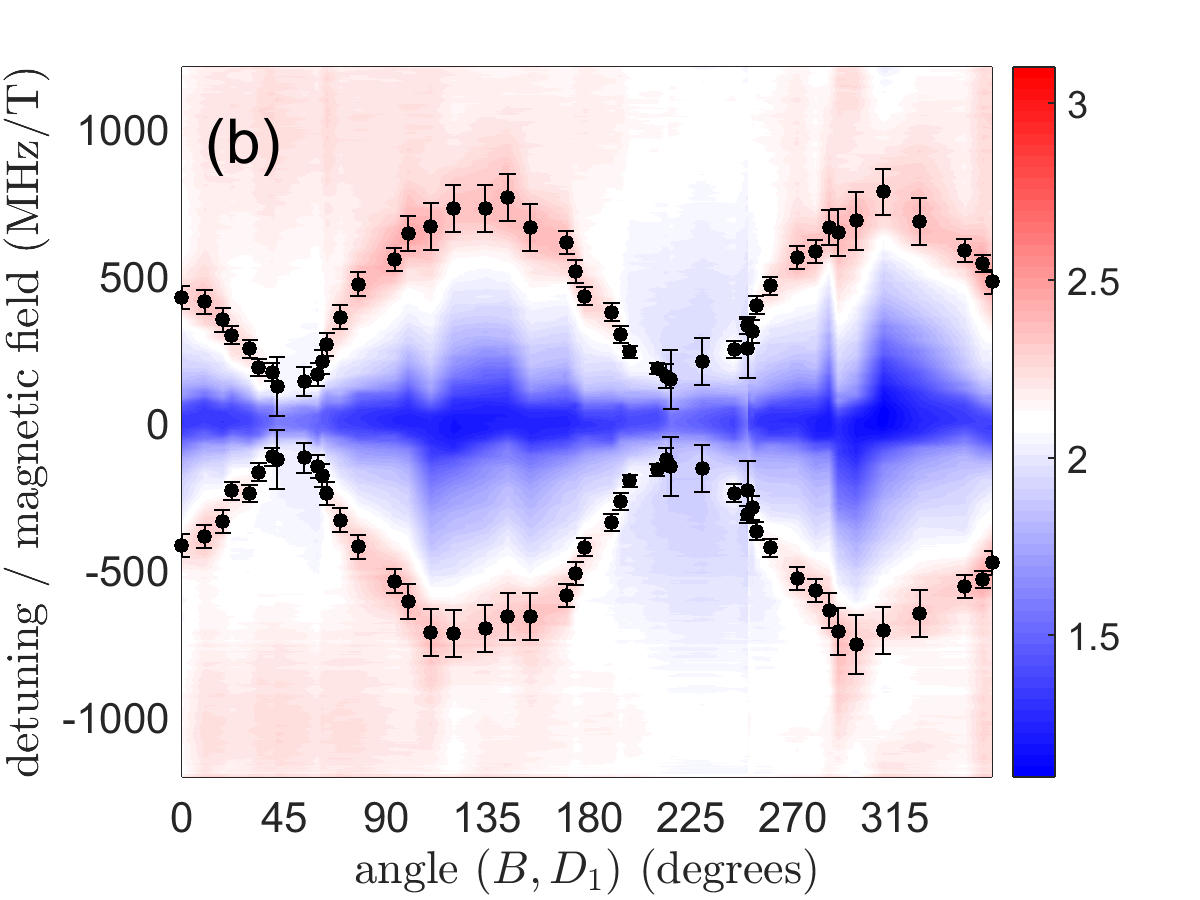}
\caption{Hole burning spectrograms with (a) $B=170$~mT and (b) $B=6.6$~mT, the magnetic field being rotated in the $(D_1,D_2)$ plane. The colorscale represents the value of the optical depth $\alpha L$. The positions of the side holes and antiholes are determined manually (without any fitting procedure) and drawn in the plot as a guide to the eye (open and closed circles, respectively). The errorbars correspond to the uncertainty of the manual determination.}
\label{fig:TmYSO_SplittingsDgDe}
\end{figure}

For the excited state we explore 25 magnetic field orientations with a $B=170$~mT magnetic field and record the corresponding holeburning spectra. We plot these spectra together as a color plot in Fig.~\ref{fig:TmYSO_SplittingsDgDe}(a). The side holes are observed with positions varying between $18$ and $42$ MHz/T.
For the ground state we explore 47 orientations with a $B=6.6$~mT magnetic field and plot the various spectra as a color plot in Fig.~\ref{fig:TmYSO_SplittingsDgDe}(b). The side antiholes are found at magnetic sensitivities comprised between $120$~MHz/T and $800$~MHz/T.

The splittings are one to two orders of magnitude larger than the pure nuclear Zeeman effect which, according to the $^{169}$Tm nuclear magnetic moment $\mu=-0.2316~\mu_N$, yields $3.5$~MHz/T splittings in the ground and excited states. Therefore we evidence an enhanced nuclear Zeeman interaction in both electronic states.

\begin{figure}[t]
\centering\includegraphics[width=0.99\linewidth]{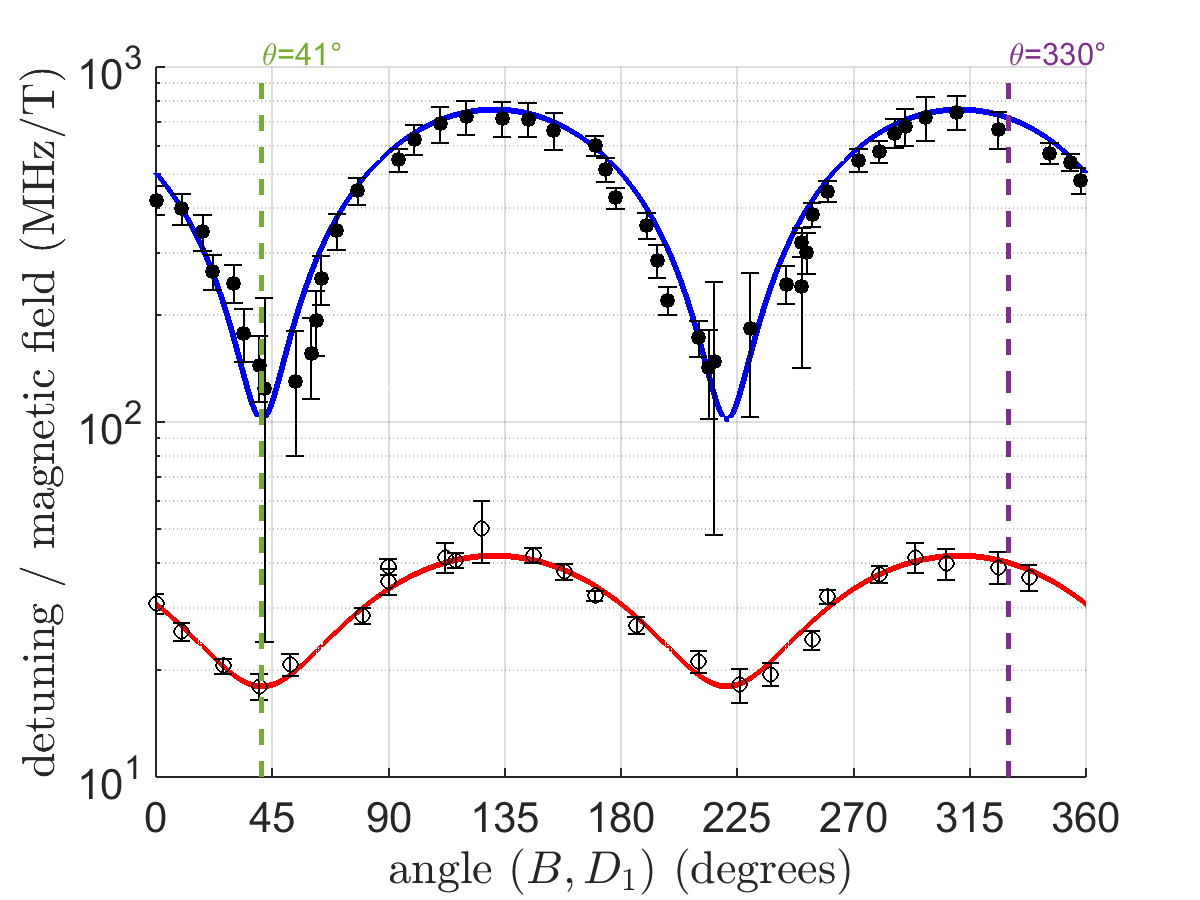}
\caption{Position sensitivity of side holes (open circles) and antiholes (closed circles). The red and blue lines represent a visual fit to the data using Eq.~(\ref{eq:Deltage_theta}).}
\label{fig:TmYSO_SplittingsGE}
\end{figure}

The enhanced nuclear Zeeman interaction is characterized by its gyromagnetic tensor $\overline{\overline{\gamma}}$, so that the splittings in ground and excited states read as:
\begin{equation}
    \Delta_i= \left\| \overline{\overline{\gamma}}_i\cdot \vec{B} \right\|
\end{equation}
where $i\in\{g,e\}$ and $\vec{B}$ is the applied magnetic field vector with $\|\vec{B}\|=B$. The tensor's principal axes are not known, and they are not necessarily aligned with the crystal axes~\cite{sun2008magnetic}.
Complete gyromagnetic tensor characterization is beyond the scope of this paper. Nevertheless, we can derive effective gyromagnetic factors in the $(D_1,D_2)$ plane, by writing:
\begin{equation}
    \Delta_i=\sqrt{\gamma_{i_x}^2 B_x^2 + \gamma_{i_y}^2 B_y^2}
    \label{eq:Deltage_D1D2}
\end{equation}
where $x$ and $y$ refer to a projection of the gyromagnetic tensor basis in the $(D_1,D_2)$ plane. This projection is characterized by the angle $\theta_0$ between the $x$ axis and $D_1$, so that the previous equation now reads as:
\begin{equation}
    \Delta_i=B\sqrt{\gamma_{i_x}^2 \cos(\theta-\theta_0)^2 + \gamma_{i_y}^2 \sin(\theta-\theta_0)^2}
    \label{eq:Deltage_theta}
\end{equation}

We locate manually the hole and antihole positions and fit their angular behavior using Eq.~\ref{eq:Deltage_theta} (see Fig.~\ref{fig:TmYSO_SplittingsGE}). The reference angle $\theta_0=41^\circ$ derived from the fit is the same for both states. The antihole experimental data exhibit a slight asymmetry, indicating that the magnetic field does not lie exactly in the crystal rotation plane.
We finally obtain the following effective gyromagnetic coefficients in the $(D_1,D_2)$ plane:
\begin{eqnarray}
      \gamma_{g_x}=120~\textrm{MHz/T}, & \hspace{0.8cm}& \gamma_{g_y}=801~\textrm{MHz/T}, \\
      \gamma_{e_x}=18~\textrm{MHz/T}, & & \gamma_{e_y}=42~\textrm{MHz/T}.
\end{eqnarray}

These effective gyromagnetic coefficients have the same order of magnitude as the ones measured in Tm:YAG~\cite{deseze2006}. The anisotropy is perhaps a little weak ($\gamma_y/\gamma_x\simeq7$ and $2$ in the ground and excited states, respectively, to be compared to $20$ and $5$ for Tm:YAG), but this could simply be explained by the possibility that none of the principal axes of the gyromagnetic tensor lie in the $(D_1,D_2)$ plane.

\subsection{Hole lifetimes}
With our $5$~ms long pumping step, significantly longer than the metastable level lifetime, we perform several optical pumping cycles and accumulate atoms in the non-resonant Zeeman ground level. We therefore expect a double exponential decay of the hole, with a fast component corresponding to the metastable state decay, and a slow decay related to population equalization in the two Zeeman sublevels of the ground states, described by the lifetime $T_Z$.

Given the gyromagnetic anisotropy in Tm:YSO, we expect the ground level lifetime $T_Z$ to be strongly anisotropic. We focus on two orientations where the magnetic sensitivity of the ground state is close to its minimum or maximum, that is $\theta=41^\circ$ and $\theta=330^\circ$, corresponding to $\theta-\theta_0=0^\circ$ and $\theta-\theta_0=289^\circ$, respectively. In both cases we set the magnetic field such that the antihole lies $\sim5$~MHz away from the central hole. This value is chosen to be of the order of the ultrasound frequencies for imaging in biological tissues (between $2$ and $15$~MHz). For comparison we also measure the hole decay with $\theta=41^\circ$ but with an inadequate field value.

\begin{figure}[t]
\centering\includegraphics[width=0.9\linewidth]{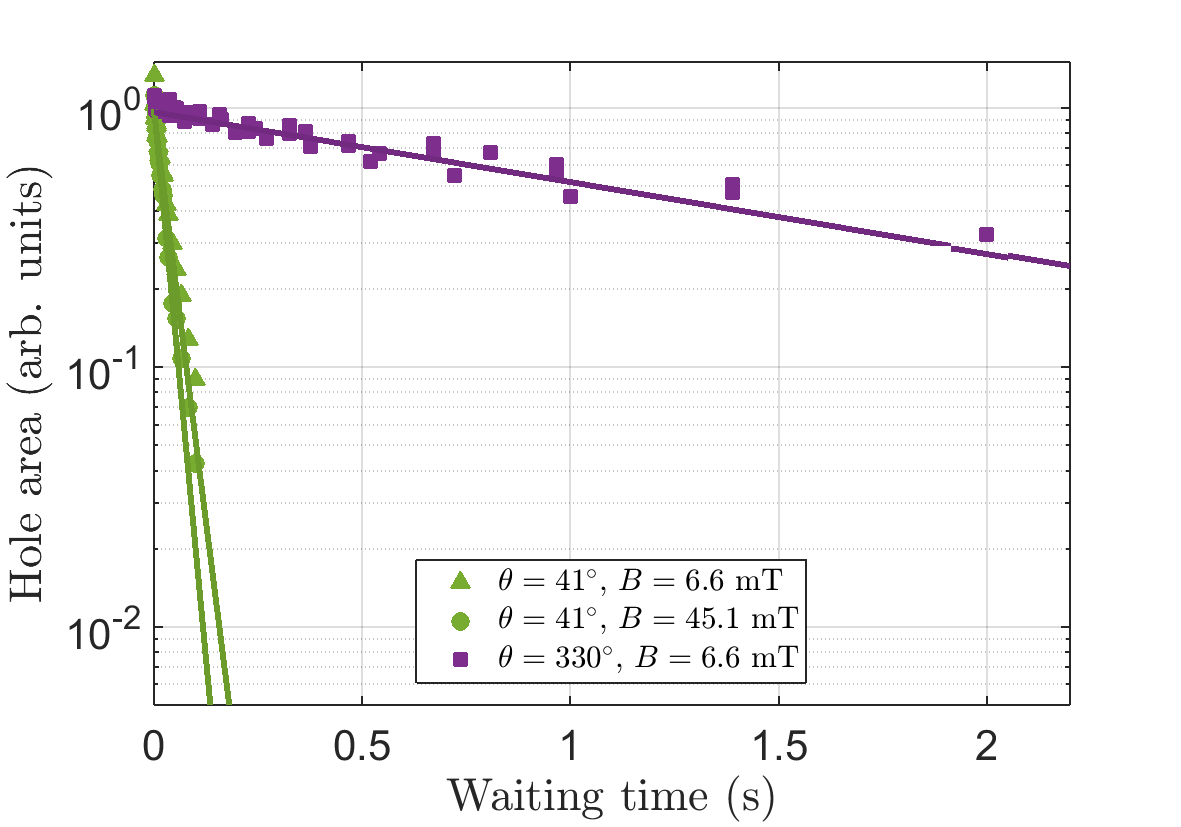}
\caption{Spectral hole decay for 2 magnetic field orientations in the $(D_1,D_2)$ plane. The symbols correspond to experimental data, and the lines to the exponential decay fit.}
\label{fig:TmYSOLifetimes}
\end{figure}

The spectral hole area evolution for waiting times between $2$~ms and $2$~s is shown in Fig.~\ref{fig:TmYSOLifetimes} and fitted with a single exponential decay $e^{-t/T_Z}$. The measured values of $T_Z$ are summed up in Table~\ref{tab:TZ}. The combination of $B=6.6$~mT and large magnetic sensitivity ($\theta=330^\circ$) leads to a persistent spectral hole, with a lifetime exceeding one second ($1.6\pm 0.25$~s). In the other magnetic field orientation, the Zeeman effect is $7$ times smaller, and intermediate storage times are observed (between $25$ and $35$~ms). The spin-lattice relaxation processes and in particular the direct process should in principle lead to a larger Zeeman lifetime for smaller magnetic sensitivity~\cite{orbach1961slr}. The inverse behaviour observed here indicates the presence of a relaxation mechanism which is frozen out by the moderate applied magnetic field, in a similar way as what was observed for example in Nd:YVO$_4$~\cite{hastings2008}.

\begin{table}[t]
\begin{tabular}{c c c c c}
\hline
$\theta$ ($^\circ$)  & $B$(mT) & $\Delta_g$(MHz)  & $\Delta_g/B$(MHz/T) & $T_Z$(ms)  \\
\hline
$41$ & $6.6$ & $0.7$ & 120 & $34\pm1.5$ \\
$41$ & $45.1$ & $4.6$ & 120 & $26\pm1$ \\
$330$ & $6.6$ & $5.0$ & 700 & $1612\pm250$ \\
\hline
\end{tabular}
\caption{Hole lifetimes in Tm:YSO under some magnetic field orientations and magnitudes at $2.1$~K. These data correspond to the spectral hole decays plotted in Fig.~\ref{fig:TmYSOLifetimes}.}
\label{tab:TZ}
\end{table}

\subsection{Burning deep holes in Tm:YSO}
\label{sec:deep}
The spectral holes shown in Fig.~\ref{fig:TrouTmYSO} are obtained with a very small laser power and therefore the residual absorption is quite substantial ($\geq 80\%$). For ultrasound tomography, deep holes are necessary to ensure large filtering dynamics.
Persistent spectral holes do not always guarantee that it is possible to burn deep holes. Indeed, the depth of a hole also depends on how efficiently one can transfer atoms from their initial state into the storage state. In the case where the storage state is reached by relaxation, the relaxation branching parameters are essential. 

In Tm-doped crystals, the relaxation from the excited state follows two paths: one direct path to the ground state, and one indirect path via the levels $^3H_5$ and $^3F_4$. In Tm:YAG for example, the transfer between Zeeman states via optical pumping is quite efficient due to the prevalence of the indirect relaxation path. The long lifetime  ($\sim10$~ms) of the metastable state $^3F_4$ then ensures a good mixing of the atomic states~\cite{ahlefeldt2015}.
In Tm:YSO, measuring these branching parameters is not an easy task because the excited state lifetime ($55~\mu$s) is shorter than the usual hole-burning sequence. Therefore, in this work, we directly demonstrate the burning of deep spectral holes.

We use a different sample holder with a fixed orientation, so that the magnetic field is now oriented with $\theta=300^\circ$ with respect to $D_1$, which is still close to the maximum Zeeman sensitivity. The magnetic field is adjusted so that the antiholes are $4.4$~MHz from the central hole. We verify the persistency of the hole in this configuration and obtain $1.36$~s hole lifetime. We choose a $200~\mu$s waiting time, and vary the burning pulse duration over almost orders of magnitude.
The incoming light power is $920~\mu$W and the beam radius is $250~\mu$m in the crystal.

\begin{figure}[t]
\centering\includegraphics[width=1.0\linewidth]{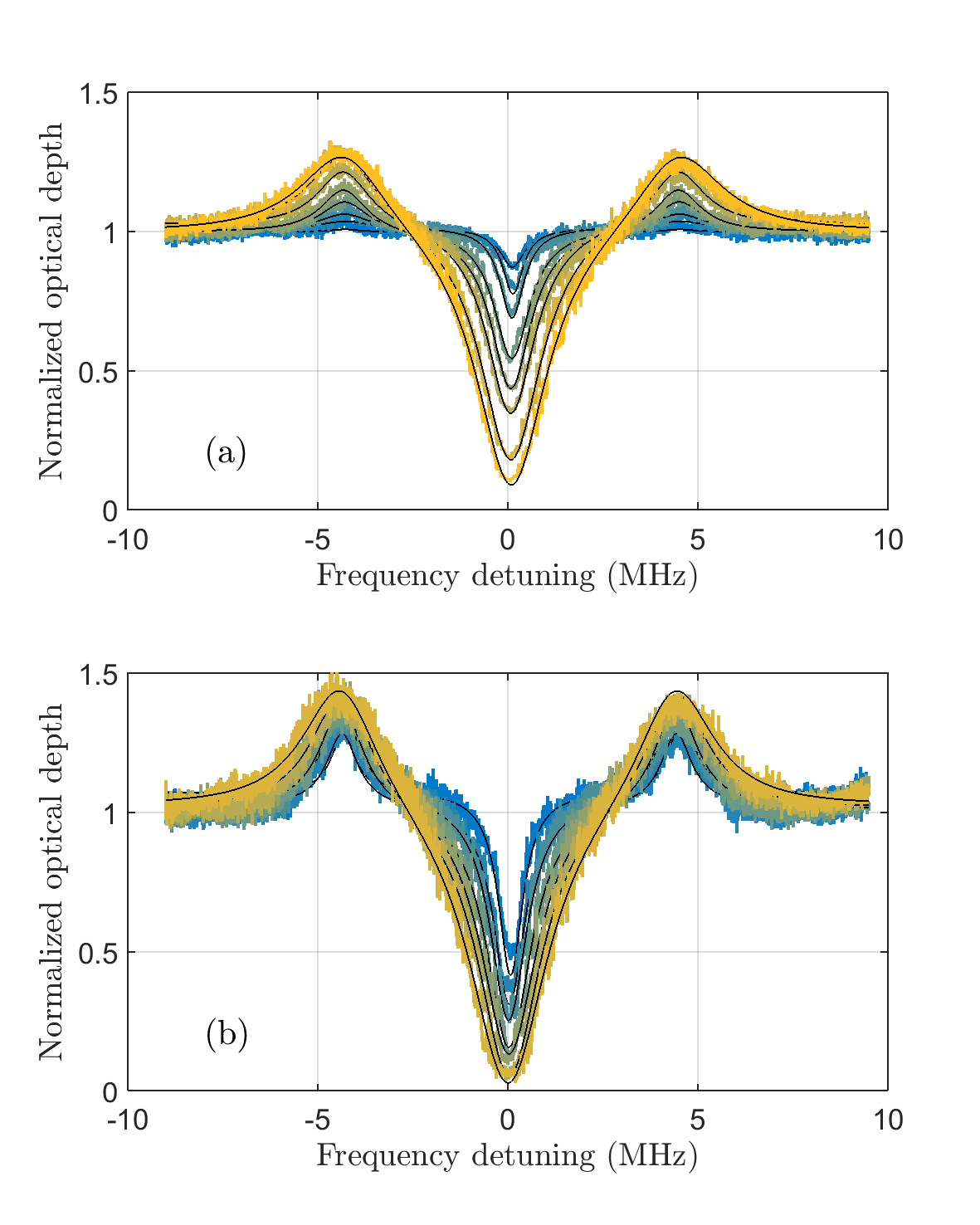}
\caption{(a) Spectral holes with burning pulse durations comprised between $50~\mu$s and $10$~ms. (b) Spectral holes with  burning pulse durations comprised between $50~\mu$s and $5$~ms, repeated every $100$~ms. In both graphs the experimental data are fitted with a sum of 3 Lorentzian functions (black lines). The optical depth is normalized with respect to the initial optical depth $\alpha L=1.65$. }
\label{fig:PumpingTmYSO}
\end{figure}

\begin{figure}[t]
\centering\includegraphics[width=1.0\linewidth]{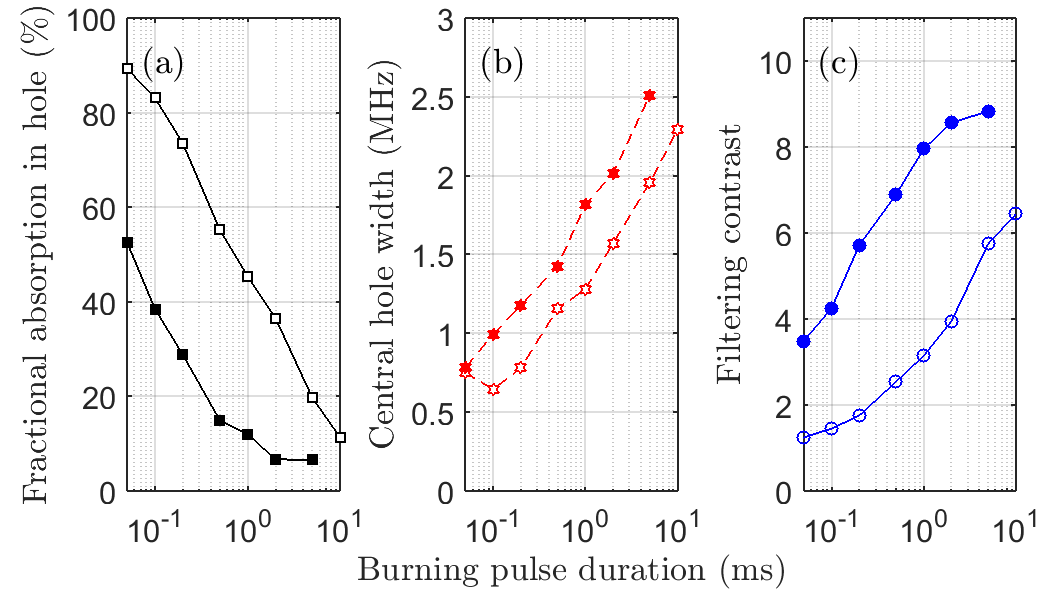}
\caption{Comparison of the spectral hole characteristics with respect to burning pulse duration and accumulation. (a): Fractional absorption at the bottom of the central hole. (b): Width of the central hole, derived from the Lorentzian fits in Fig~\ref{fig:PumpingTmYSO}. (c): Filtering contrast between the center hole and a side antihole, using $\alpha L=1.65$ as initial optical depth. In all three graphs, the open symbols correspond to the experiment using a single burning pulse, whereas the closed symbols correspond to the periodically repeated burning pulses.}
\label{fig:PumpingTmYSO_resume}
\end{figure}

Figure~\ref{fig:PumpingTmYSO}(a) shows the resulting spectral holes in the absorption profile. This profile is normalized with respect to the optical depth $\alpha_0 L=1.65$, the laser being slightly detuned from the absorption peak for this experiment. As expected, the hole depth steadily increases with the burning pulse duration [see Fig.~\ref{fig:PumpingTmYSO_resume}(a)], until the residual absorption at the center is only $10$\% of the initial absorption. The hole width also increases with the burning duration due to saturation, from $700$~kHz up to $2.2$~MHz [see Fig.~\ref{fig:PumpingTmYSO_resume}(b)]. The position of the antiholes $\pm 5$~MHz from the central hole actually limits this width to half the distance of the antiholes, \emph{ie} $2.5$~MHz, due to competing pumping and depumping processes in the two ground Zeeman sublevels.

\subsection{Towards ultrasound optical tomography in Tm:YSO}
We have demonstrated deep, long-lived, sub-MHz spectral holes in Tm:YSO under magnetic field. This is very promising for implementing a spectral filter in a ultrasound optical tomography (UOT) setup. However, realistic UOT conditions require a quasi-stationary spectral hole shape over the acquisition time which can last several minutes while still allowing for a large duty cycle dedicated to imaging. UOT also requires a large pumping beam waist (of the order of $1$~cm$^{2}$) to be compatible with large optical \emph{\'etendue}, and a large filtering dynamic range, ideally above $30$~dB. 

To fully appreciate the potential of Tm:YSO as a spectral filter for UOT, we perform additional holeburning experiments, similar to those presented in Sec.~\ref{sec:deep} but with periodically repeated burning pulses every $100$~ms. This way, the hole is refreshed when it has lost only  $7\%$ of its maximum size and can be considered as stationary. As shown in Figures~\ref{fig:PumpingTmYSO}(b) and \ref{fig:PumpingTmYSO_resume}, the repetition of the burning pulses leads to deeper and wider spectral holes thanks to accumulated optical pumping. In particular, we reach as little as $6\%$ residual absorption at the bottom of the central hole with repeated $5$~ms pulses. In this configuration, only $5\%$ of the duty cycle is occupied by the maintenance of the spectral hole, leaving the remaining $95$\% for imaging.

\begin{figure}[t!]
\centering\includegraphics[width=0.9\linewidth]{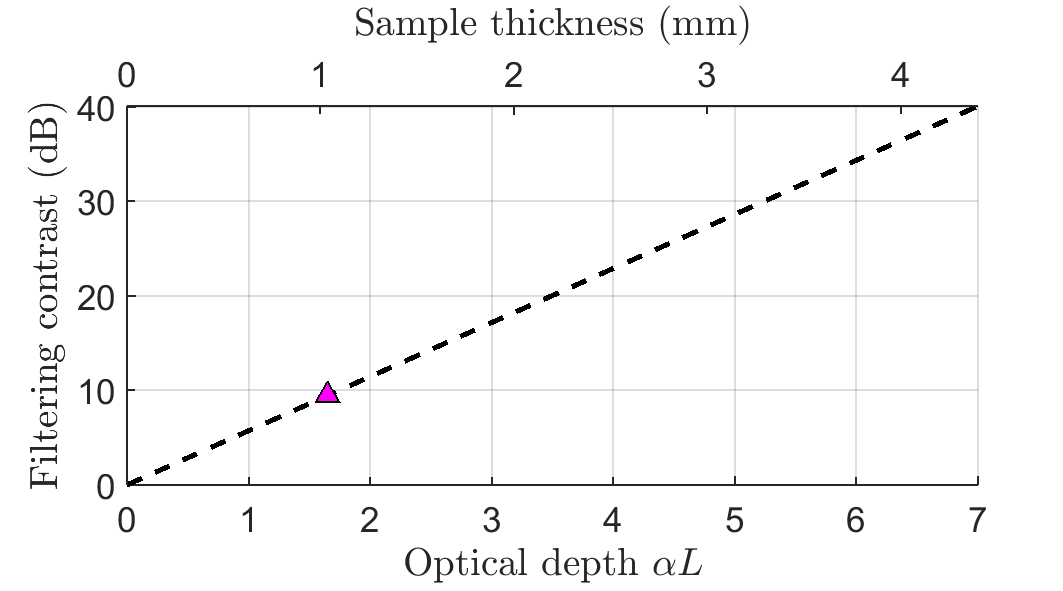}
\caption{Extrapolation of the filtering contrast, expressed in dB, as a function of the optical depth. The corresponding sample thickness is also shown in the top $x$-axis, considering a $0.1$\%-doped Tm:YSO sample and considering the laser tuned to the center of the inhomogeneous absorption line. The triangle shows the maximal experimental filtering contrast demonstrated in Fig.~\ref{fig:PumpingTmYSO}(b).}
\label{fig:Extrapolation} 
\end{figure}

In a standard UOT experiment, the signal of interest is tuned to the minimum of absorption (the center of the hole). The carrier light (a few MHz away), on the other hand, undergoes the nominal absorption~\cite{li2008uot}. As already shown in~\cite{venet2018}, the dynamic range of the filter can be enhanced if the carrier light coincides with one of the antiholes where the absorption is enhanced. We define the filtering contrast as the ratio between the crystal transmission at the center of the main hole and the crystal transmission at the position of the antiholes. 
This filtering contrast is equal to $8.8$ for the present value of $\alpha L=1.65$, with repeated $5$~ms pulses. This number can be easily increased by tuning the laser to the center of the line, and also by using a longer crystal or a larger doping concentration. We extrapolate the filtering contrasts for larger values of the crystal optical depth in Fig.~\ref{fig:Extrapolation}, neglecting the absorption of the burning pulses through the crystal and assuming a uniform spectral hole over the crystal thickness. This assumption is supported by the small value of the residual absorption in the main hole. With a $0.1\%$-doped Tm:YSO sample three times longer ($6$~mm) and with a laser tuned exactly on resonance, we calculate a nominal optical depth of $\alpha L=6.6$ and expect a filtering contrast around $6000$. The emblematic $30$~dB limit should be reached with $\alpha L\simeq5.5$.

Regarding the spatial extension of the spectral hole, the pumping rate is proportional to the optical irradiance expressed in W/cm$^2$. Extending the beam size up to $1$~cm in diameter (by a factor of $20$) leads to a dramatic drop of the irradiance which can be compensated by increasing the laser power by a factor of $400$, \emph{ie} $370$~mW. Such a laser power value is accessible with standard tapered amplifiers.

Although extrapolated and based on simple assumptions, these orders of magnitude show the promising potential of Tm:YSO as a spectral filter for UOT, with fully manageable laser power and crystal size.


\section{Conclusion}
We have investigated Tm-doped YSO under magnetic field via holeburning spectroscopy. We have found the input polarization configuration maximizing the absorption. We have identified a transient holeburning mechanism lasting $740~\mu$s, that we assign to storage in the metastable state ${}^3F_4$. By applying an external magnetic field, we have observed the anisotropic enhanced nuclear Zeeman effect and partly characterized the gyromagnetic tensor coefficients in the ground and excited states in the $(D_1,D_2)$ plane of the YSO host.
Although our study is far from exhaustive, we have found a magnetic field configuration where Tm:YSO exhibits persistent spectral hole burning, with hole lifetimes longer than a second, and succeeded in burning deep spectral holes. Using our experimental results with a limited crystal optical depth we have estimated the experimental parameters that would correspond to a real-scale ultrasound optical tomography setup, requiring a large hole contrast together with a wide beam radius, and have established that these parameters are accessible.
Combined with the optical therapeutic window wavelength compatibility, the large absorption coefficient, the narrow inhomogeneous optical line, and the sub-MHz width of the spectral holes, these results confirm Tm:YSO as a promising compound for this application.

\begin{acknowledgments}
The authors acknowledge fruitful discussions with Thierry Chaneli{\`e}re and Philippe Goldner, as well as financial support from the MALT project (grant C16027HS) from ITMO Cancer AVIESAN (Alliance Nationale pour les Sciences de la Vie et de la Sant\'e, National Alliance for Life Sciences \& Health) within the framework of the Cancer Plan.
\end{acknowledgments}

\bibliography{TmYSO}

\end{document}